\documentclass[aps,twocolumn,nofootinbib,groupedaddress]{revtex4-1}
\usepackage{amssymb,amsmath,amstext,amsfonts}
\usepackage{bbold}
\usepackage{bm}
\usepackage{graphicx}
\usepackage{xcolor}

\newcommand{\beq}{\begin{equation}}
\newcommand{\eeq}{\end{equation}}
\newcommand{\bal}{\begin{aligned}}
\newcommand{\eal}{\end{aligned}}

\newcommand{\Mp}{M_{\rm Pl}}
\newcommand{\va}{{\varphi}}
\newcommand{\M}{M}
\newcommand{\fNLsquashed}{f_{NL}^{\rm flat}}
\newcommand{\Pad}{{\cal P}_0}

\begin{document}

\title{Hyper non-Gaussianities in inflation with strongly non-geodesic motion} 

\author{Jacopo Fumagalli, Sebastian Garcia-Saenz, Lucas Pinol, S\'ebastien Renaux-Petel and John Ronayne}
\affiliation{Institut d'Astrophysique de Paris, GReCO, UMR 7095 du CNRS et de Sorbonne Universit\'e,
98bis boulevard Arago, Paris 75014, France}

\date{\today}

\begin{abstract}

Several recent proposals to embed inflation into high-energy physics rely on inflationary dynamics characterized by a strongly non-geodesic motion in negatively curved field space. This naturally leads to a transient instability of perturbations on sub-Hubble scales, and to their exponential amplification. Supported by first-principle numerical computations, and by the analytical insight provided by the effective field theory of inflation, we show that the bispectrum is enhanced in flattened configurations, and we argue that an analogous result holds for all higher-order correlation functions. These ``hyper non-Gaussianities'' thus provide powerful model-independent constraints on non-standard inflationary attractors motivated by the search for ultraviolet completions of inflation.

\end{abstract}

\maketitle


{\bf Introduction.}--- Negatively curved field space plays a crucial role in modern embeddings of inflation in high-energy physics. Non-linear sigma models with a hyperbolic target space arise naturally in top-down realizations of inflation, particularly within supergravity, giving rise to the $\alpha$-attractor class of models (see \textit{e.g.}~\cite{Kallosh:2013yoa,PhysRevD.92.041301,Achucarro:2017ing}). Independently of the question of their ultraviolet completions, non-minimal kinetic terms of the hyperbolic type lead to interesting dynamics, allowing for non-trivial inflationary trajectories characterized by a strongly non-geodesic motion \cite{Achucarro:2016fby,Christodoulidis:2018qdw,Garcia-Saenz:2018ifx,Bjorkmo:2019fls}.  This in turn relaxes the conditions of slow-roll to allow for potentials that are steep in Planck units \cite{Hetz:2016ics,Achucarro:2018vey}, a welcome feature in view of the eta problem and the recently much discussed swampland conjectures \cite{Obied:2018sgi,Agrawal:2018own}. Lastly, internal field spaces with negative curvature are at the origin of the phenomenon of geometrical destabilization \cite{Renaux-Petel:2015mga,Renaux-Petel:2017dia,Krajewski:2018moi,Cicoli:2018ccr,Grocholski:2019mot}, in which non-inflationary degrees of freedom, even heavy ones, can dramatically affect the fate of inflation. 

A concrete scenario in which the consequences of a hyperbolic field space have been studied is the proposal of ``hyperinflation''  \cite{Brown:2017osf}, that has recently been under scrutiny \cite{Mizuno:2017idt,Bjorkmo:2019aev}. The intuitive picture of this set-up is that of an inflationary trajectory corresponding to a circular motion around the minimum of a (circularly symmetric) scalar potential. The hyperbolic geometry is crucial to compensate for the loss of angular velocity to the Hubble friction, allowing inflation to last long enough, even if the potential is too steep to inflate along a radial trajectory. Within this circumstance, hyperinflation proceeds along a strongly non-geodesic trajectory, and a striking outcome is an exponential growth of the curvature power spectrum around the time of Hubble crossing, and the corresponding suppression of the tensor-to-scalar ratio. With such an amplification, assessing the size of nonlinear effects in this setup appears to be crucial, while previous studies have restricted their attention to the analysis of linear fluctuations.

In this context, this Letter presents a general framework to study non-Gaussianities in the presence of strongly non-geodesic motion typical of hyperbolic-type geometry, highlighting how this naturally leads to \textit{``hyper non-Gaussianities''}. For definiteness we concentrate on the specific example of hyperinflation as a particularly interesting playground to analyze the effects of the non-trivial field space in this class of models. However our results are formulated in general terms and have a broad range of applicability. Essentially, they indicate that in negatively curved field space, inflationary models with strongly non-geodesic motion are characterized by an enhanced non-Gaussian signal, both for the bispectrum and for all higher-point correlation functions, that can easily lead to tensions with experimental bounds. These model-independent constraints sharpen the range of allowed theoretical constructions, and are of utmost importance in view of the intense current efforts to build non-standard inflationary scenarios in agreement with quantum gravity conjectures.\\


{\bf Hyperinflation.}--- The starting point is an action for two scalar fields $\va^I=(\phi,\theta)$ with non-canonical kinetic term minimally coupled to gravity:
\beq \label{eq:action1}
S=\int d^4x\sqrt{-g}\bigg[\frac{\Mp^2}{2}\,R-\frac{1}{2}\,G_{IJ}\nabla^{\mu}\va^I\nabla_{\mu}\va^J-V(\va)\bigg]\,.
\eeq
The matrix $G_{IJ}$ defines a metric in the internal field space parametrized by the coordinates $\va^I$, in this case the hyperbolic plane of curvature $-2/\M^2$, and is assumed to have the form
\beq
G_{IJ}d\va^Id\va^J=d\phi^2+\M^2\sinh^2\left(\frac{\phi}{\M}\right)d\theta^2\,.
\eeq
Moreover, the potential is assumed to depend only on the ``radial'' field $\phi$, $V=V(\phi)$, with $V'>0$.

Consider now an inflationary background characterized by homogeneous fields $\phi(t)$ and $\theta(t)$, and a quasi-de Sitter spacetime metric with scale factor $a(t)$ and Hubble parameter $H(t)=\dot{a}/a$, with $t$ the cosmological time.
Hyperinflation corresponds to a non-standard attractor solution of the action \eqref{eq:action1}, with small parameters $\epsilon\equiv-\dot{H}/H^2 \simeq 3 \M V'/2V$ and $\eta \equiv \dot{\epsilon}/H\epsilon \simeq 2 \epsilon -3 \M V''/V'$, that arises under the conditions 
\beq
 \frac{3 \M}{\Mp} < \frac{\Mp V'}{V} \ll \frac{\Mp}{\M}\,,\qquad \frac{\M |V''|}{V'}\ll1\,.
 \label{conditions}
\eeq
More precisely, the equation of motion $\ddot{\phi}+3H\dot{\phi}-\M\sinh\left(\phi/\M\right)\cosh\left(\phi/\M\right)\dot{\theta}^2+V'(\phi)=0$
admits a solution with $\dot{\phi}\simeq -3 \M H$, independently of the slope of the potential. Defining $h^2\equiv \frac{V'(\phi)}{\M H^2}-9$, a positive quantity for hyperinflation solutions (see \eqref{conditions}), one has $h^2/9+1 \simeq \epsilon_V/\epsilon \simeq \eta_V/(2\epsilon-\eta)$, where $\epsilon_V =\frac12\Mp^2 (V'/V)^2$ and $\eta_V=\Mp V''/V$ are the standard potential ``slow-roll'' parameters, not necessarily small here. In hyperinflation, potentials that verify the swampland de Sitter conjecture (in its refined version \cite{Garg:2018reu,Ooguri:2018wrx}) should obey either $\epsilon_V \geq {\cal O}(1)$ or $-\eta_V \geq {\cal O}(1)$, corresponding respectively to a steep slope or steep negative curvature in Planck units. From the above relations, one deduces that a prolonged phase of hyperinflation supported by such potentials is necessarily characterized by $h^2 \gg 1$ (see also \cite{Bjorkmo:2019aev}). We will concentrate on this theoretically most interesting regime, which, as we will see, corresponds to a strongly non-geodesic motion. Actually, as emphasized recently in a model-independent manner \cite{Achucarro:2018vey}, the latter feature is necessary in order to inflate on potentials whose slope is steep in Planck units, and it is also a characteristic feature of the sidetracked models studied in Ref.~\cite{Garcia-Saenz:2018ifx}.


{\bf Strongly non-geodesic motion and dynamics of linear perturbations.}--- Let us now consider linear fluctuations. We employ gauge invariant variables $Q^I$ that coincide with the field fluctuations $\delta\va^I$ in the spatially flat gauge, and perform a decomposition in terms of the adiabatic and entropic modes $Q_{\sigma}$ and $Q_s$, defined as the projection of $Q^I$ in the direction tangential and perpendicular to the background trajectory respectively, the inner product being defined by the field space metric $G_{IJ}$ (see \textit{e.g.}~\cite{Gong:2016qmq} for a review of perturbation theory in multifield inflation). The adiabatic mode can be expressed as $Q_{\sigma}=\frac{\dot{\sigma}}{H}\,\zeta$, with the definition $\dot{\sigma}\equiv\sqrt{G_{IJ}\dot{\va}^I\dot{\va}^J}$, and where $\zeta$ is the comoving curvature perturbation. The equation of motion for the entropic perturbation is given by
\beq \label{eq:Qs eom}
\ddot{Q}_s+3H\dot{Q}_s+\left(\frac{k^2}{a^2}+m_s^2\right)Q_s=-2\dot{\sigma}\eta_{\perp}\dot{\zeta}\,,
\eeq
where $k$ is the Fourier wavenumber, we introduced 
\beq
\eta_\perp\equiv -\frac{V_{,s}}{H\dot{\sigma}}\,,\qquad m_s^2\equiv V_{;ss}-H^2\eta_{\perp}^2+ \epsilon\, H^2 R_{{\rm fs}} \Mp^2\,.
\label{ms}
\eeq
Here $V_{,s}$ and $V_{;ss}$ stand for the projections in the entropic direction of the first and second (field-space covariant) derivatives of the potential, respectively, and $R_{{\rm fs}}$ is the field space scalar curvature. The ``bending'' parameter $\eta_\perp$ is physically important as it gives a measure of the deviation of the background trajectory from a geodesic in field space \cite{GrootNibbelink:2001qt}. A strongly non-geodesic motion is characterized by $\eta_\perp^2 \gg 1$, resulting in a large negative contribution to the entropic mass $m_s^2$, something that a negatively curved field space only reinforces. Without a stabilization from the potential, a large negative $m_s^2/H^2$ is thus a built-in feature of these models. Although unusual, this property is not \textit{a priori} in contradiction with the requirement of a stable background. Indeed, on super-Hubble scales $k/a\ll H$ one has $\dot{\zeta}=2H^2\eta_\perp/\dot{\sigma}\,Q_s$, and \eqref{eq:Qs eom} yields an uncoupled equation for $Q_s$, now with a different effective mass $m_{s({\rm eff})}^2\equiv m_s^2+4H^2\eta_{\perp}^2$, to which the bending contributes positively.

Specifying this general discussion to hyperinflation, we find, to leading-order in the slow-varying approximation,
\beq
\eta_{\perp}^2\simeq h^2\,,\qquad m_s^2\simeq -2H^2h^2\,, \qquad m_{s({\rm eff})}^2\simeq 2H^2h^2\,.
\eeq
As anticipated, a strongly non-geodesic motion corresponds to $h^2 \gg 1$. In this situation, the large and {\it positive} $m_{s({\rm eff})}^2/H^2$ implies a rapid decay of $Q_s$ on super-Hubble scales and the conservation relation $\dot{\zeta}\simeq0$. The decay of homogeneous perturbations $Q_s(k=0)$ is a proof that the background solution of hyperinflation is indeed a stable attractor. On the contrary, the large and {\it negative} $m_s^2/H^2$ signals a transient instability of fluctuations on sub-Hubble scales, observed also in some sidetracked inflationary models \cite{Garcia-Saenz:2018ifx} and earlier in Refs.~\cite{Cremonini:2010ua,Kaiser:2012ak}.
\begin{figure}
  \includegraphics*[width=8cm]{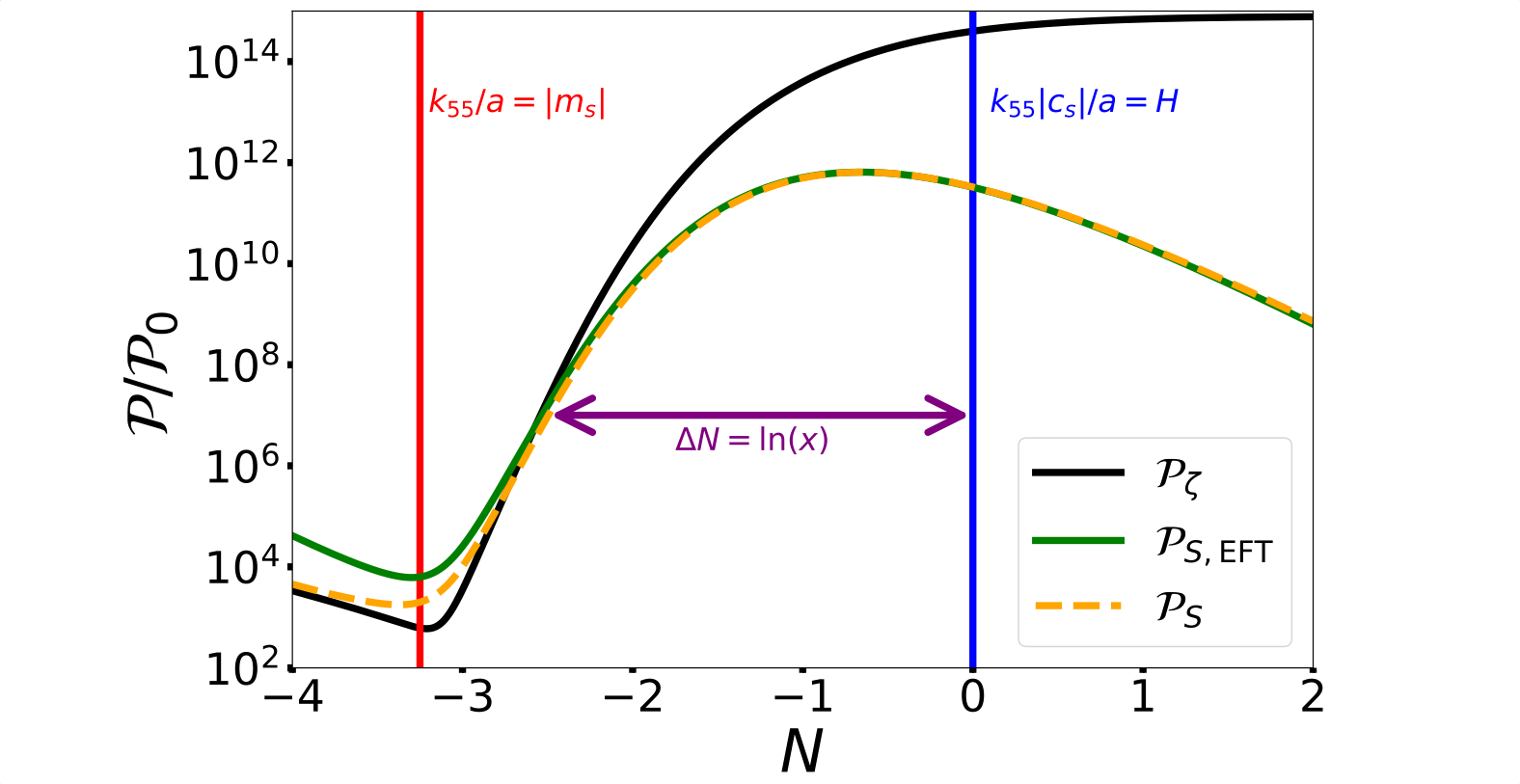}
\caption{Adiabatic (${\cal P}_{\zeta}$) and entropic (${\cal P}_{{\cal S}}$) power spectra as functions of the number of $e$-folds, for the representative model detailed in the main text. The spectra are evaluated for the scale $k_{55}$ that crosses the Hubble radius $55$ $e$-folds before the end of inflation, at $N=0$ in the plot. ${\cal P}_{{\cal S},{\rm EFT}}$ is the entropic power spectrum deduced from the relation \eqref{Qs-integrating-out}.
\label{fig:power-spectrum}}
 \end{figure}
These considerations can be checked numerically. In Fig.~\eqref{fig:power-spectrum} we consider the same background as studied in \cite{Mizuno:2017idt}, for the quadratic potential $V(\phi)=\frac{1}{2}m^2\phi^2$ with $m=\M=10^{-2}\Mp$. We display the time-dependence of the dimensionless curvature (${\cal P}_\zeta$) and entropic (${\cal P}_{{\cal S}}=H^2/{\dot \sigma}^2 {\cal P}_{Q_s}$) power spectra, for the scale $k_{55}$ that crosses the Hubble radius 55 $e$-folds before the end of inflation, which we take as the CMB pivot scale. The normalization factor is the ``standard'' result $\Pad=H^2/(8 \pi^2 \epsilon \Mp^2)_{k_{55}=aH}$. Soon after ``entropic mass crossing'' the exponential growth of entropic fluctuations caused by the tachyonic instability feeds the adiabatic perturbation, before entropic fluctuations decay and the curvature perturbation becomes constant. The transient instability results in an exponentially small tensor-to-scalar ratio $r= 3.6\times10^{-16}$. 

\begin{figure}
\hspace{-0.5cm}
  \includegraphics*[width=9cm]{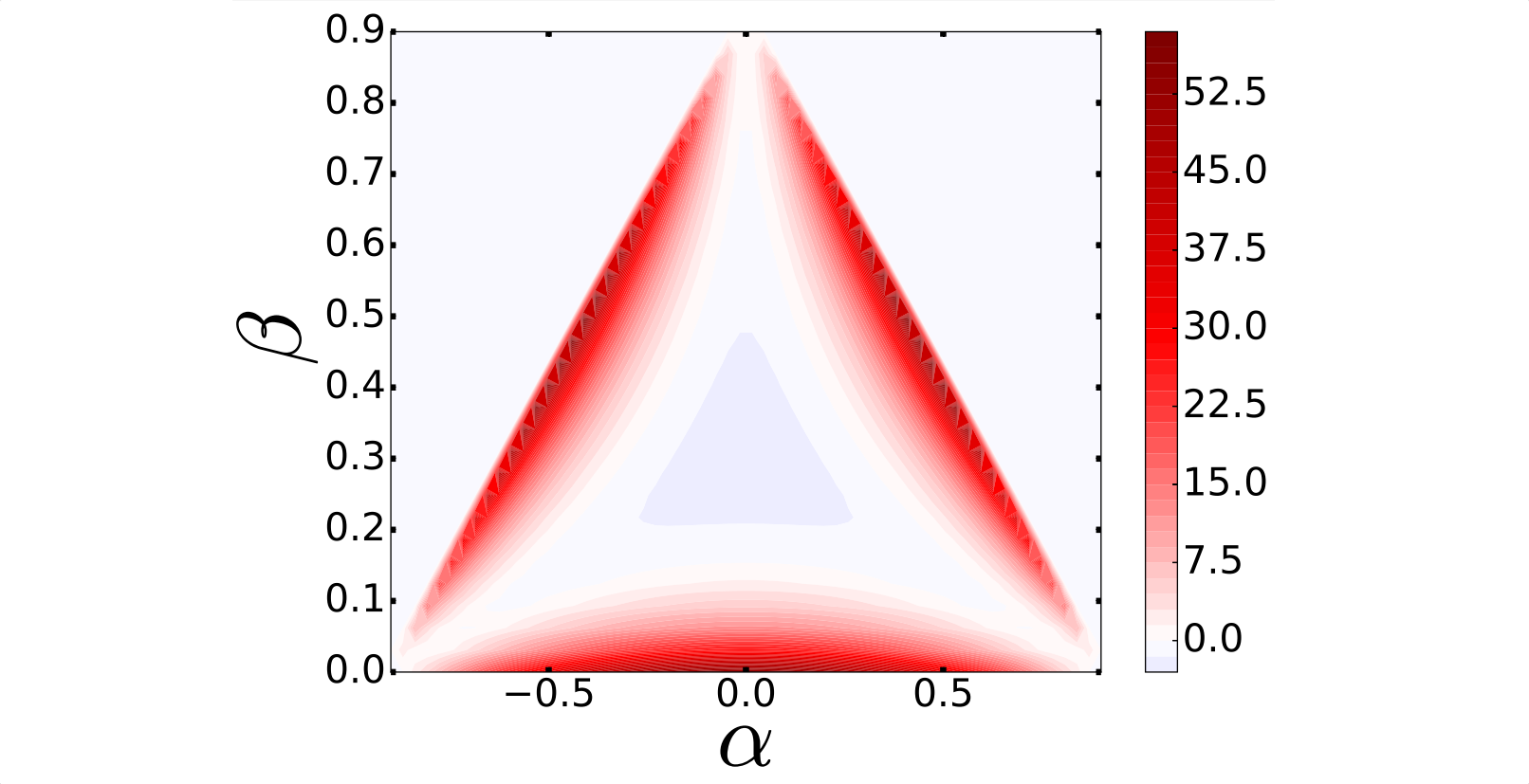}
\caption{Shape dependence $f_{NL}(\alpha,\beta)$, at fixed overall scale $k_1+k_2+k_3=3 k_{55}$, for the same model as in Fig.~\ref{fig:power-spectrum}. The characteristic feature of the bispectrum is its dominant signal near flattened configurations. Note that the equilateral configuration corresponds to the point $(\alpha,\beta)=(0,1/3)$. 
 \label{fig:fNL}}
 \end{figure}

 
{\bf Bispectrum.}--- We now turn to the numerical calculation of the bispectrum in hyperinflation.\footnote{See \textit{e.g.}~\cite{Wang:2013eqj,Renaux-Petel:2015bja} for recent reviews on primordial non-Gaussianities, and \cite{Akrami:2019izv} for the most recent observational constraints from the Planck collaboration.} The results have been obtained with {\texttt{PyTransport 2.0}} \cite{Mulryne:2016mzv,Ronayne:2017qzn}, a code based on the transport approach to compute two- and three-point correlation functions in multifield models with curved field space (see also {\texttt{CppTransport}} \cite{Dias:2016rjq,Seery:2016lko,Butchers:2018hds}). Fig.\ \ref{fig:fNL} is a plot of the reduced bispectrum $f_{NL}(k_1,k_2,k_3)$ for the same representative model as above, as a function of the variables $(\alpha,\beta)$ defined by $k_1=\frac{3k_{55}}{4}(1+\alpha+\beta)$, $k_2=\frac{3k_{55}}{4}(1-\alpha+\beta)$ and $k_3=\frac{3k_{55}}{2}(1-\beta)$. The resulting bispectrum is quite unlike what is usually found in inflationary models with Bunch-Davies vacuum state. In particular, hyperinflation generates a non-Gaussian signal that is peaked near flattened triangle configurations, \textit{i.e.}~the ones with $k_1\simeq k_2+k_3$ (the edges in Fig.\ \ref{fig:fNL}), which is typical of excited initial states. Explicitly, we find 
\beq \label{eq:fNL numerical}
f_{NL}^{\rm eq}=-2.0\,,\qquad \fNLsquashed=53.8\,,
\eeq
where the two parameters simply denote the evaluation of the reduced bispectrum, respectively at $k_1=k_2=k_3$ and at the representative flattened configuration $k_2=k_3=k_1/2$.
Although we have presented results for the particular case of a quadratic potential and for a specific set of parameters, we remark that the qualitative outcome --- a strong non-Gaussian signal in flattened configurations --- is quite robust. We find for instance $\fNLsquashed \simeq25$ for a Starobinsky-type potential and $\fNLsquashed \simeq100$ for a quartic potential, while $f_{NL}^{\rm eq}=O(1)$ in each case.


{\bf Effective single-field description.}--- In a standard setup with a single light degree of freedom, heavy entropic fluctuations can be integrated out to yield an EFT for the adiabatic mode (see \textit{e.g.}~\cite{Tolley:2009fg,Cremonini:2010ua,Achucarro:2010da,Achucarro:2012sm,Achucarro:2012yr}). In the type of models we consider, the entropic field is heavy but {\it tachyonic}, yet the procedure can be carried out equivalently to the standard case, as explained in \cite{Garcia-Saenz:2018vqf}. In slowly evolving backgrounds and in the regime $k^2/a^2\ll|m_s^2|$, one finds
\beq
Q_s=-2 \frac{\dot{\sigma}\eta_{\perp}}{m_s^2} \dot{\zeta}\,,
\label{Qs-integrating-out}
\eeq
which results in an effective quadratic action for the curvature perturbation:
\beq \label{eq:quadratic eft}
S^{(2)}_{\rm eff}=\int d\tau d^3x\,a^2\epsilon \Mp^2 \left[\frac{\zeta^{\prime2}}{c_s^2}-(\vec{\nabla}\zeta)^2\right]\,,
\eeq
where $\tau \simeq -1/(aH)$ is the conformal time, $\zeta'\equiv d\zeta/d\tau$, and the speed of sound $c_s$ is defined by
\beq \label{eq:definition of c_s}
\frac{1}{c_s^2}\equiv 1+\frac{4H^2\eta_{\perp}^2}{m_s^2}=\frac{m_{s({\rm eff})}^2}{m_s^2}\,.
\eeq
An imaginary sound speed is thus a model-independent consequence of a tachyonic entropic mass ($m_s^2<0$) and a stable background ($m_{s({\rm eff})}^2>0$). For instance in hyperinflation we find $c_s^2\simeq -1\,.$ In this class of models, the curvature perturbation thus propagates with a ``wrong sign'' dispersion relation: $\omega^2=-|c_s|^2k^2\simeq -k^2$. Accordingly, the mode functions do not oscillate but rather grow or decay exponentially, affecting all wave modes $k$ up to the cutoff of the EFT --- what in the two-field theory was a tachyonic instability affecting low-energy modes has become in the EFT a gradient-type instability to which the whole spectrum is sensitive. Such a theory is not a priori catastrophic, as the corresponding instability is only transient. However, our results will ultimately bound the amplification of fluctuations.

It will prove useful to introduce the dimensionless parameter $x$ such that the wave modes $k$ described by the effective theory satisfy $k|c_s|/a<xH$. Its dependence on models' parameters is $x \sim |c_s| |m_s|/H $, where the numerical factor in the right should be somewhat smaller than unity. Quantitatively, $x$ can be determined by examining when the power spectrum of the entropic fluctuation computed in the full theory matches the one deduced from \eqref{Qs-integrating-out}, which is the central relation from which the EFT derives. One can see in Fig.~\ref{fig:power-spectrum} that the two become in very good agreement less than one $e$-fold after entropic mass crossing, corresponding to $x \sim 10$ in this example. The general solution of the linear equation of motion reads, for $c_s^2<0$,
\begin{eqnarray}
\zeta_k(\tau)&=&\left(\frac{2 \pi^2}{k^3}\right)^{1/2}\alpha \left( e^{k |c_s| \tau+x}(k |c_s| \tau-1) \right.
  \nonumber\\
  &&\hspace*{+1.0em}-\left. \rho e^{i \psi} e^{-(k |c_s| \tau+x)}(k |c_s| \tau+1) \right)\,,
\label{mode-function}
\end{eqnarray}
where we omit the mild $k$-dependence of $(\alpha,\rho,\psi)$ for simplicity, and we stress that it only applies for $k|c_s|\tau+x \geq 0$. The parameter $\rho$ sets the relative amplitude of the exponentially decaying mode compared to the growing one at the time marking the validity of the EFT (and $\psi$ is a phase difference), while $\alpha$ can be taken to be real and parameterizes their overall amplitude.\footnote{The amplitude of the decaying mode is necessarily non-zero, as the quantization condition entails the relation $2 \alpha^2 \rho \sin(\psi) |c_s| =\Pad$.} The final value of the curvature power spectrum ${\cal P}_{\zeta}= \alpha^2\,e^{2x}$ (assuming very conservatively that $\rho \lesssim {\cal O}(1)$) depends on the initial conditions of the EFT. Although these can in principle be determined by matching to the full computation of ${\cal P}_{\zeta}(\tau)$, interestingly, this is not needed to study higher-order correlation functions, to which we now turn.

The cubic action of the EFT of inflationary perturbations \cite{Creminelli:2006xe,Cheung:2007st} (at lowest order in derivatives and in the slow-varying approximation):
\beq\bal
S^{(3)}_{\rm eff}&=\int d\tau d^3x\,\frac{a\, \epsilon\Mp^2}{H}\left(\frac{1}{c_s^2}-1\right) \left[\zeta'(\vec{\nabla}\zeta)^2+\frac{A}{c_s^2}\,\zeta^{\prime3}\right]\,,
\label{S3}
\eal\eeq
where $A$ is a dimensionless constant of order $1$, that can be in principle computed from the full theory. Although the interactions in \eqref{S3} are standard, the behavior of the mode function \eqref{mode-function} is not, and the computation of the bispectrum is nontrivial \cite{Garcia-Saenz:2018vqf}. Contrary to what a naive power counting would indicate, one finds that the reduced bispectrum does not feature an exponential enhancement by $e^{2x}$, like the power spectrum. Instead, it is independent of $x$ for near equilateral configurations, with
\beq \label{eq:fNLeq EFT}
f_{NL}^{\rm eq}\simeq\frac{10}{9}\left(\frac{1}{|c_s|^2}+1\right)\left(\frac{13A}{6}-\frac{5}{24}\right)\,.
\eeq
Given the result $c_s^2\simeq-1$ in hyperinflation, we have the analytical prediction that $f_{NL}^{\rm eq}=O(1)$, in agreement with the numerical results. Away from the equilateral limit, one finds a dependence of $f_{NL}$ on the cutoff scale $xH$ (with the exception of the squeezed limit, which can be shown to verify the single-clock consistency relation via standard arguments). Akin to models with excited initial states (see \textit{e.g.}~\cite{Chen:2006nt,Holman:2007na,Meerburg:2009ys,Meerburg:2009fi,Agarwal:2012mq}) the constructive interferences between two growing and one decaying mode result in a magnification of the signal near flattened configurations $k_1\simeq k_2+k_3$, 
\beq\bal \label{eq:fNLflat EFT}
\fNLsquashed&\simeq\frac{1}{192}\left(\frac{1}{|c_s|^2}+1\right)\\
&\quad\times\left(39(A-1)+12x^2+4(A+1)x^3\right)\,,
\eal\eeq 
and a global shape with a large overlap with the orthogonal template (except for values of $A \simeq -1$) \cite{Garcia-Saenz:2018vqf}. With $x^2 \sim h^2 \gg 1$ in hyperinflation with strongly non-geodesic motion (\textit{e.g.}~$x^2 \sim 100$ in our example), we conclude that flattened non-Gaussianities are large in this type of model, again in agreement with the full two-field numerical results (see the Supplemental Material for a quantitative comparison).


{\bf Higher-order correlation functions.}--- We have seen that the single-field effective theory of perturbations with imaginary sound speed can unambiguously predict the striking features of the reduced bispectrum, thus providing a valuable device to gain analytical insight into the complex multifield dynamics of models with large negative entropic mass. But from a pragmatic perspective, the real power of the EFT approach is that it can go beyond the reach of current numerical methods, as we now show by providing an estimate of higher-order correlation functions in this class, including hyperinflation. Like for the bispectrum, for $n \geq 4$, the reduced connected $n$-point function $\langle\zeta^n\rangle/\langle \zeta^2\rangle^{n-1}$ does not feature an exponential amplification \cite{Bjorkmo:2019qno}.\footnote{This has been shown in \cite{Bjorkmo:2019qno} using the EFT put forward in the first arXiv version of this work.}
In spite of this, we find that some flattened configurations for the trispectrum and higher-point correlators are enhanced by powers of the parameter $x$, essentially due to the same phenomenon of constructive interferences between growing and decaying modes that occurs for the bispectrum.

\begin{figure}
\vspace{0.6cm}
\hspace{-0.7cm}
  \includegraphics*[width=6.6cm]{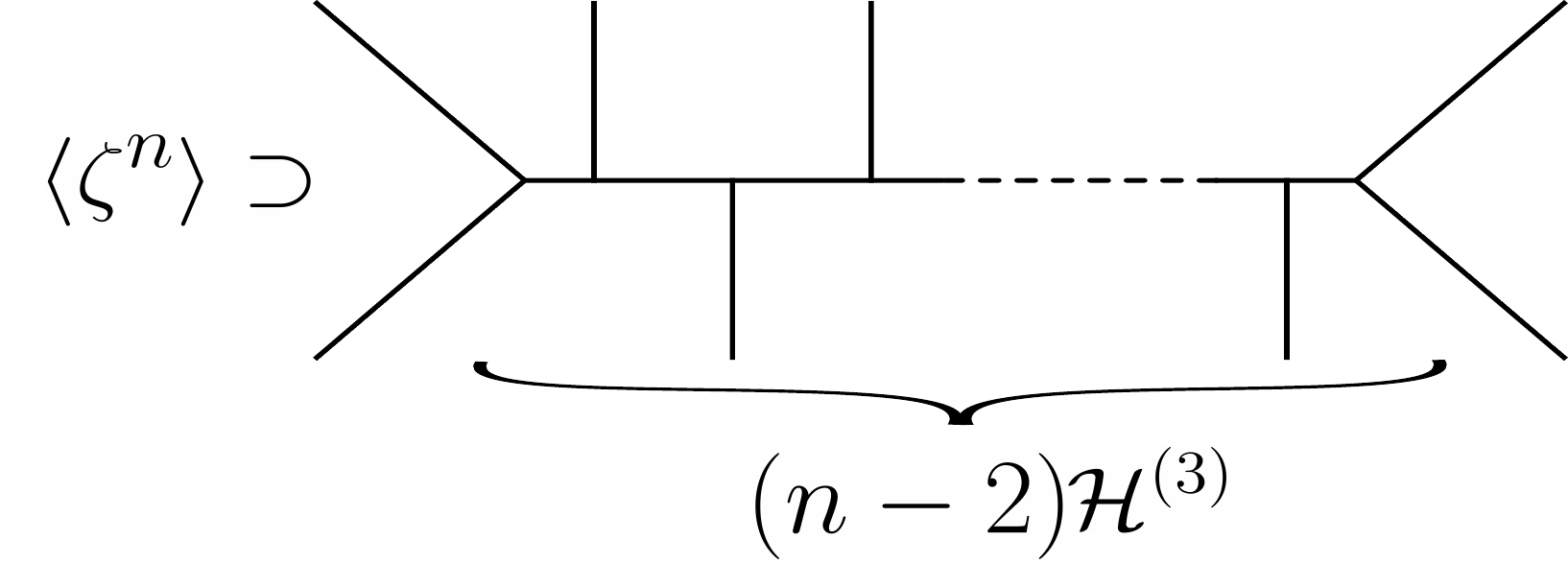}
\caption{Dominant contribution to the connected $n$-point correlation function of $\zeta$.
\label{fig:n-point}}
\end{figure}

Focusing on the dominant contributions in the large $x$ limit, one can find an explicit derivation of the trispectrum in the Supplemental Material, where we also give further details of our estimate for the higher-point functions and characterize which flattened shapes are enhanced. 
Following the \textit{in-in} formalism \cite{Weinberg:2005vy}, the computation of the $n$-point correlator can be organized as a sum of connected Feynman diagrams with $n$ external lines, and with each insertion of the interaction Hamiltonian corresponding to a vertex. 
We find that a diagram with $v$ vertices contributes as $\langle\zeta^n\rangle/\langle \zeta^2\rangle^{n-1}\propto x^{2n+v-4}$ for the maximally enhanced flattened configurations. This contribution is largest when $v=n-2$, corresponding to diagrams of the type shown in Fig.\ \ref{fig:n-point}, with $(n-2)$ insertions of the cubic Hamiltonian, yielding the estimate
\beq \label{eq:n-pt function}
\frac{\langle\zeta^n\rangle}{\langle \zeta^2\rangle^{n-1}}\sim \left[ \left( \frac{1}{|c_s|^2}+1\right) x^3\right]^{n-2}\,.
\eeq
Notice that, despite the rapid growth of the correlation functions as $n$ increases, 
the theory is nevertheless under perturbative control for observationally relevant models.
Indeed, perturbativity is guaranteed provided
$\langle\zeta^n\rangle/\langle \zeta^2\rangle^{n-1} \times {\cal P}_{\zeta}^{(n-2)/2} \sim \left( \fNLsquashed \,A_s^{1/2}\right)^{n-2}\lesssim 1$,
which is a weaker requirement than meeting the observational bounds on the bispectrum.


{\bf Discussion.}---  In negatively curved field space, and in the absence of a stabilizing effect from the potential in the direction perpendicular to the background trajectory, a strongly non-geodesic motion in field space automatically induces a transient instability of fluctuations on sub-Hubble scales. Under these general circumstances, we can make use of an effective field theory for the curvature perturbation that naturally explains the exponential amplification of the power spectrum. Moreover, it predicts a reduced bispectrum whose characteristics are in striking agreement with first-principles numerical computations in the full theory: order-one non-Gaussianities in equilateral configurations, and a magnification near flattened ones. However, contrary to the power spectrum, the reduced bispectrum is not exponentially amplified. We have moreover argued that an analogous outcome holds for all higher-order reduced correlation functions, namely a hierarchical enhancement for particular flattened shapes proportional to a power of the instability rate. When the latter is very large, these \textit{``hyper non-Gaussianities''} lead to tensions with observational constraints, as exemplified by models of hyperinflation that satisfy the de Sitter swampland conjecture.

Our model-independent results severely bound the magnitude of a large negative entropic mass.
Hence, it results in a powerful selection criterion on models with negatively curved field space and a strongly non-geodesic motion, that have been receiving much attention recently. Namely, at least some stabilizing effect by the potential is needed to counterbalance the otherwise strongly tachyonic mass of entropic fluctuations on sub-Hubble scales. This compensating effect can be large, to the extent that $|m_s^2/H^2| \ll 1$ or $m_s^2/H^2 \gg 1$, which are well understood situations. It can also be mild, resulting again in a negative entropic mass, but not parametrically larger than the Hubble scale (a feature shared by hyperinflation with a moderate degree of bending). Without hierarchy, an EFT cannot be rigorously derived, but we expect our results to give a qualitatively correct picture, \textit{i.e.}~an enhancement of the bispectrum in flattened configurations, and similarly for the trispectrum. Thus, our results pave the way for future studies about the role of field-space geometry in the dynamics of inflation, and particularly in how non-geodesic motion of inflationary attractors can lead to novel signatures that can be probed with current and next-generation experiments.


\begin{acknowledgments}

We are very grateful to D.\ Andriot, T.\ Bjorkmo, P.\ Christodoulidis, A.\ Joyce, D.\ Marsh, S.\ Mizuno, S.\ Mukohyama, A.\ Nicolis, E.\ Pajer, D.\ Roest and K.\ Turzy\'nski for interesting and helpful discussions. J.F, L.P, S.RP and J.R are supported by the European Research Council under the European Union's Horizon 2020 research and innovation programme (grant agreement No 758792, project GEODESI). SGS is supported by the European Research Council under the European Community's Seventh Framework Programme (FP7/2007-2013 Grant Agreement no.\ 307934, NIRG project); he would also like to thank the Van Swinderen Institute for Particle Physics and Gravity for generous hospitality.

\end{acknowledgments}


\bibliographystyle{apsrev4-1}
\bibliography{arXiv-v2.bbl}


\vspace{3.5cm}
{\Large {\bf Supplemental Material}}\\

\subsection{Comparison between analytical and numerical bispectra}

In this first supplemental material we provide further details on the comparison of the results for the bispectrum as computed numerically in the full two-field theory and analytically in the single-field EFT. In the main text we have quoted the reduced bispectrum $f_{NL}$ in the equilateral and squashed configurations (the latter being representative of more general flattened configurations): Eq.~\eqref{eq:fNL numerical} gives the numerical results (for the specific set of parameters $m=\M=10^{-2}\Mp$) while Eqs.~\eqref{eq:fNLeq EFT} and \eqref{eq:fNLflat EFT} give the approximate analytical formulae as derived from the EFT.

The EFT formulae depend on $c_s$, $A$ and $x$, which are unknown parameters within the EFT and must be determined either from the knowledge of the UV completion (the hyperinflation model in this case) or through a matching calculation. The speed of sound is easily computed in the two-field theory from the standard result \eqref{eq:definition of c_s}, which gives $c_s^2\simeq-1$ in hyperinflation, as indicated in the main text. For the constants $A$ and $x$ it is easier to perform a matching given the knowledge of the full numerical bispectrum. Remembering that the EFT result for $f_{NL}$ is more accurate for the equilateral configuration \cite{Garcia-Saenz:2018vqf}, we compare $f_{NL}^{\rm eq}$ in Eqs.~\eqref{eq:fNL numerical} and \eqref{eq:fNLeq EFT} to find
\beq \label{eq:A numerical estimate}
A\simeq -0.33\,.
\eeq
On the other hand, the computation of $x$ is necessarily less accurate, since $x$ is precisely what defines the regime of validity of the EFT. In the text we showed how the value of $x$ may be estimated from the entropic power spectrum, which produced the result $x\sim10$.\\

\begin{figure}
\includegraphics*[width=7.5cm]{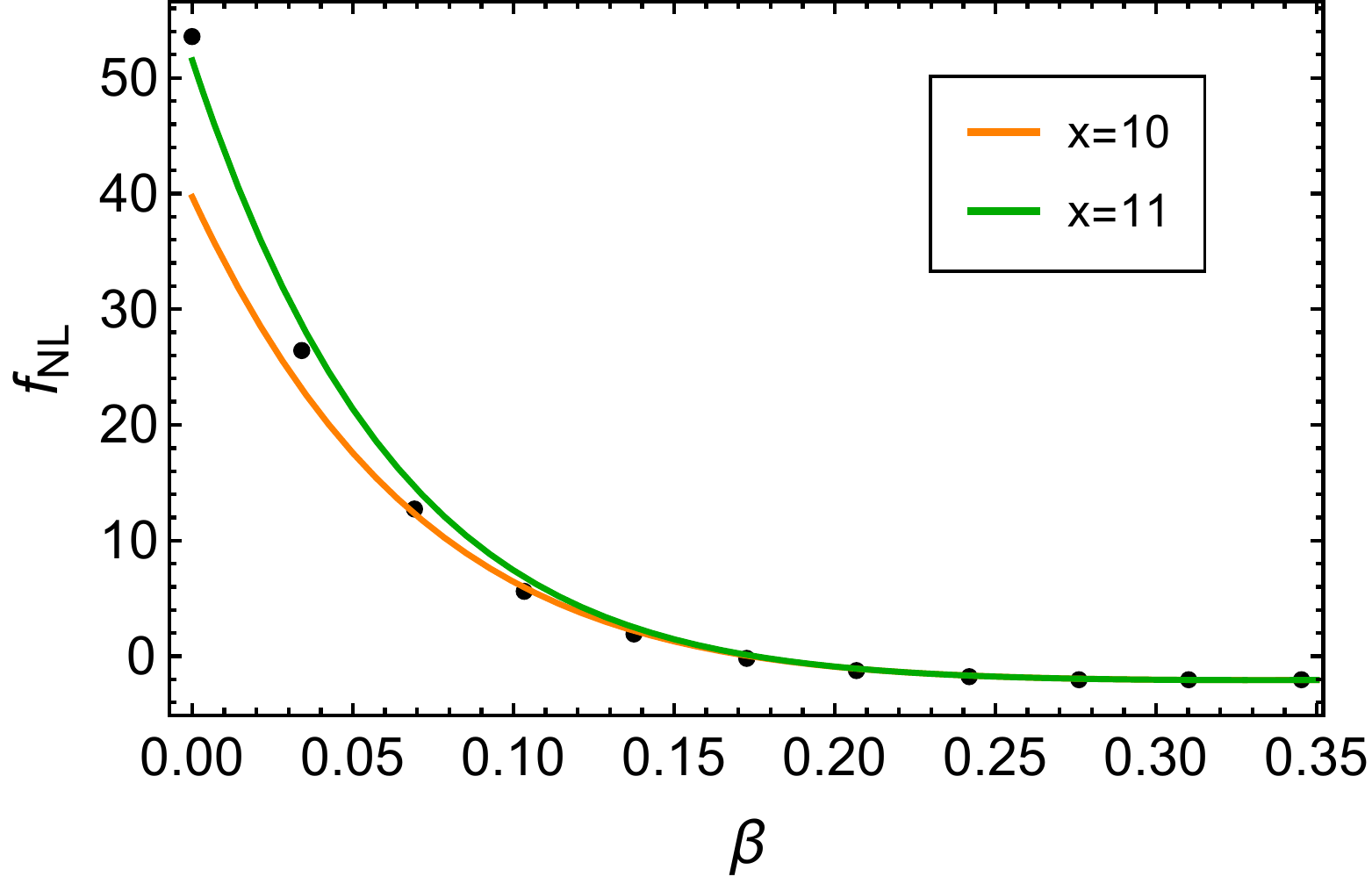}
\caption{Graph of the function $f_{NL}(0,\beta)$ corresponding to a one-dimensional slice of the reduced bispectrum. The points are the numerical data, and the curves show the analytical EFT results for $c_s^2=-1$, $A=-0.33$ and two different values of $x$.}
\label{fig:fNL slice}
\end{figure}

We now make this estimate for $x$ more precise by comparing the full analytical and numerical bispectra. Fig.~\ref{fig:fNL slice} displays the one-dimensional slice of $f_{NL}(\alpha,\beta)$ that ranges from $f_{NL}(0,0)\equiv f_{NL}^{\rm flat}$ to $f_{NL}(0,1/3)\equiv f_{NL}^{\rm eq}$, where for the EFT prediction we have used $x=10$ and $x=11$. It is clear that the estimate derived from the 2-point statistics provides a quantitatively good match for the bispectrum. 
As expected, the dependence on $x$ is only important near the flattened configuration, and we judge the value $x\simeq10$ as a slightly better fit to the numerical data in view of the fact that the EFT prediction for $f_{NL}^{\rm flat}$ is expected to be less accurate than for $f_{NL}^{\rm eq}$. 

\begin{figure}
\includegraphics*[width=9cm]{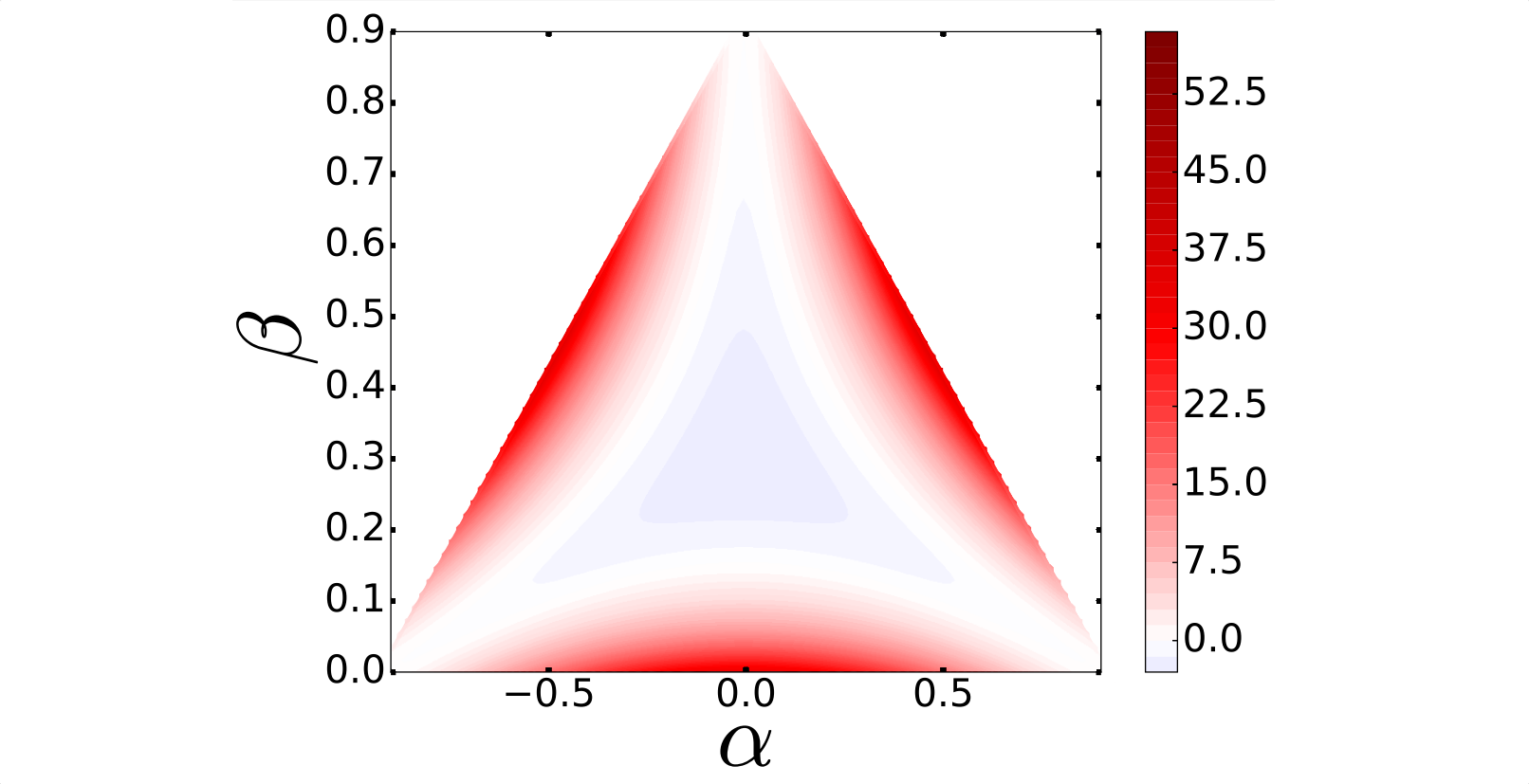}
\caption{Shape dependence $f_{NL}(\alpha,\beta)$ derived analytically in the EFT of single-field inflation, for $x=10$ and $A=-0.33$. Observe the very good agreement with the full numerical result Fig.~2 in the main text.}
 \label{fig:fNL EFT}
 \end{figure}

Fig.~\ref{fig:fNL EFT} is the full plot of the reduced bispectrum derived from the EFT, using the value $x=10$
and $A=-0.33$ as in \eqref{eq:A numerical estimate}. Although we have used a single one-dimensional slice of $f_{NL}$ for the estimation of $x$ and $A$, one can see an overall excellent agreement for the global shape of the bispectrum between the analytical and numerical results (see Fig.~2 in the main text).

Incidentally, even though the evaluation of the EFT coefficient $A$ via a matching computation is perfectly sufficient, it is worth remarking that it also concords with the analytical prediction inferred from the two-field UV completion. Applying the results of Ref.\ \cite{Garcia-Saenz:2019njm} to the model of hyperinflation, we find
\beq
A=-\frac{1}{3}-\frac{1}{6(9+h^2)}\,\frac{MV'''}{H^2}\,.
\eeq
For the simple quadratic potential that we used in this Letter we have $V'''=0$. This yields $A=-1/3$, in perfect agreement with the matching calculation.


\subsection{Estimation of the trispectrum and higher-point correlation functions}

In this second supplemental material we derive explicitly the trispectrum for the shape configurations that dominate in the limit of large $x$, focusing on a specific type of vertex for simplicity. We then generalize the calculation to estimate the reduced $n$-point function and derive Eq.~\eqref{eq:n-pt function}. For this, we use that the $n$-point correlator is given by the \textit{in-in} formula \cite{Weinberg:2005vy}
\beq\bal \label{eq:in-in formula}
\langle\zeta^n\rangle&=\sum_{j=0}^{\infty}i^j\int_{\tau_0}^0d\tau_1a(\tau_1)\int_{\tau_0}^{\tau_1}d\tau_2a(\tau_2)\cdots\int_{\tau_0}^{\tau_{j-1}}d\tau_ja(\tau_j)\\
&\quad\times \langle0|[H_I(\tau_j),\cdots[H_I(\tau_1),\zeta^n_I]\cdots]|0\rangle\,,
\eal\eeq
where the subscript $I$ refers to the interaction picture, $H_I$ is the interaction Hamiltonian, and we have introduced the ultraviolet cutoff $\tau_0=-x/(k_m|c_s|)$ to regularize the time integrals (with $k_m$ the largest among the external momenta), consistent with the fact that the EFT is only valid after a certain time.

\subsubsection{Trispectrum from $\zeta^{\prime3}$ interaction}

We begin by defining
\beq
{\cal T}(k_1,k_2,k_3,k_4,k_{12},k_{13})=\frac{\langle\zeta_{\vec{k}_1}\zeta_{\vec{k}_2}\zeta_{\vec{k}_3}\zeta_{\vec{k}_4}\rangle(k_1k_2k_3k_4)^{9/4}}{(2\pi)^9\delta(\sum_i\vec{k}_i)A_s^3}\,,
\eeq
as a dimensionless measure of the reduced 4-point correlation function. As usual, because of isotropy the trispectrum can be parametrized in terms of the magnitudes $k_i$ of the momenta plus two of the sums $k_{ij}\equiv|\vec{k}_i+\vec{k}_j|$, which we choose to be $k_{12}$ and $k_{13}$. The above formula also includes the amplitude of the scalar power spectrum at the CMB pivot scale, which in terms of our EFT parameters (and ignoring their scale dependence) is $A_s\simeq \frac{\alpha^2}{2\pi^2}\,e^{2x}$.

The trispectrum receives contributions from two types of diagrams: a contact 4-point diagram coming from the quartic interactions of the theory and a scalar exchange diagram involving two insertions of the cubic Hamiltonian in Eq.~\eqref{eq:in-in formula}. The full calculation is of course complicated by the fact that the Hamiltonian includes several different operators, so for the sake of simplicity we will focus only on the $\zeta^{\prime3}$ operator in \eqref{S3}, although we have checked that other interactions do not qualitatively modify our results. In other words, we ignore the operator $\zeta'(\vec{\nabla}\zeta)^2$ as well as the whole quartic Hamiltonian (it will be clear in a moment, however, that the latter gives a subdominant contribution), and write 
\beq
H^{(3)}=\int d^3x\,{\cal C}\,\zeta^{\prime3}\,,\qquad {\cal C}\equiv -\frac{\Mp^2\epsilon}{H}\left(\frac{1}{|c_s|^2}+1\right)\frac{A}{|c_s|^2}\,,
\eeq
and we take ${\cal C}$ to be constant to leading order in slow-roll. Inserting this in Eq.~\eqref{eq:in-in formula} and performing some standard manipulations we arrive at 
\beq\bal \label{eq:trispectrum result2}
&\langle\zeta_{\vec{k}_1}\zeta_{\vec{k}_2}\zeta_{\vec{k}_3}\zeta_{\vec{k}_4}\rangle=-144(2\pi)^3{\cal C}^2\delta({\textstyle\sum}_i\vec{k}_i)\int_{\tau_0}^0\frac{d\tau_1}{H\tau_1}\int_{\tau_0}^{\tau_1}\frac{d\tau_2}{H\tau_2}\\
&\times\Big({\rm Im}\left[\zeta'_{k_1}(\tau_2)\zeta'_{k_2}(\tau_2)\zeta'_{k_{12}}(\tau_2)\zeta^{\prime*}_{k_{12}}(\tau_1)\zeta^{*}_{k_1}(0)\zeta^{*}_{k_2}(0)\right]\\
&\quad\times{\rm Im}\left[\zeta'_{k_3}(\tau_1)\zeta'_{k_4}(\tau_1)\zeta^{*}_{k_3}(0)\zeta^{*}_{k_4}(0)\right]+(\mbox{5 perm.})\Big)\,.
\eal\eeq
This formula includes one imaginary part of product of mode functions for each of the vertices in the exchange diagram. For this reason, the decaying mode in Eq.~\eqref{mode-function} must be taken into account in some of the $\zeta$'s, and we isolate the leading term in the large $x$ regime by choosing exactly one decaying mode per vertex. Substituting we obtain an expression of the form
\beq \bal \label{eq:trispectrum result1}
&{\cal T}=\frac{9A^2|c_s|^6}{32}\left(\frac{1}{|c_s|^2}+1\right)^2(k_1k_2k_3k_4)^{5/4}\\
&\times\Bigg[k_{12}\sum_{\tilde{k}_1,\tilde{k}_2}\sigma_{\tilde{k}_1,\tilde{k}_2}\int_{\tau_0}^{0}d\tau_1\tau_1^2e^{\tilde{k}_1|c_s|\tau_1}\int_{\tau_0}^{\tau_1}d\tau_2\tau_2^2e^{\tilde{k}_2|c_s|\tau_2}\\
&\quad+(\mbox{5 perm.})\Bigg]\,,
\eal \eeq
where $\tilde{k}_1=\pm k_{12}\pm k_3\pm k_4$ and $\tilde{k}_2=\pm k_1\pm k_2\pm k_{12}$, and the sum is over all choices of signs in $\tilde{k}_1$ and $\tilde{k}_2$, but subject to the condition that exactly one of the momenta in each imaginary part in Eq.\ \eqref{eq:trispectrum result2} should carry a minus sign ($\sigma_{\tilde{k}_1,\tilde{k}_2}=\pm1$ is an overall sign that depends on the specific permutation). From this result it is clear that the configurations of momenta that maximize the absolute value of ${\cal T}$ (in terms of its scaling with $x$) are achieved when $\tilde{k}_1=\tilde{k}_2=0$, corresponding for each vertex to constructive interactions between two growing and one decaying mode.
This condition notably enforces the four external momenta to be collinear, leading to a quadrilateral shape of the flattened type. Note however that this collinearity is necessary but not sufficient; not all flattened shapes lead to a maximally enhanced trispectrum. Although analyzing the full shape-dependence of the trispectrum is beyond the scope of this Letter, let us quote the result for the configuration that is the natural extension of the representative flattened configuration maximizing the bispectrum, and evaluated in \eqref{eq:fNL numerical}. It corresponds to $\vec{k}_4=-3\vec{k}_1=-3\vec{k}_2=-3\vec{k}_3$ (so that $k_{1,2,3}=k_m/3$, $k_4=k_m$, $k_{12}=2k_m/3$) and gives
\beq
{\cal T}=-\mathcal{N}A^2\left(\frac{1}{|c_s|^2}+1\right)^2x^6\,,
\eeq
with a numerical prefactor $\mathcal{N}^{-1}=2592\times 3^{3/4}$.

\subsubsection{Higher-point functions}

Eq.\ \eqref{eq:trispectrum result1} can be generalized  to an arbitrary tree diagram with $n$ external legs. Consider such a diagram with $v$ vertices, each corresponding to an insertion of the Hamiltonian $H^{(m_j)}$, where $m_j\geq3$ and $j\in\{1,\ldots,v\}$. By dimensional analysis we have $H^{(m_j)}\sim \frac{\Mp^2\epsilon}{H^{m_j-2}a^{m_j-3}}\int d^3x(\zeta')^{m_j}$, where we emphasize that the numerical prefactor depends on $|c_s|$ and on various Wilson coefficients. It is convenient to change variables in the integrals in \eqref{eq:in-in formula} with $y_j=-k_m|c_s|\tau_j$. Then the generalization of \eqref{eq:trispectrum result1} is given by
\beq\bal
&\frac{\langle\zeta^n\rangle}{\langle \zeta^2\rangle^{n-1}}\sim \int_0^x dy_1\,y_1^{2m_1-4}e^{-(\tilde{k}_1/k_m)y_1}\\
&\times\int_{y_1}^x dy_2\,y_2^{2m_2-4}e^{-(\tilde{k}_2/k_m)y_2}\\
&\times\cdots\times\int_{y_{v-1}}^x dy_v\,y_v^{2m_v-4}e^{-(\tilde{k}_v/k_m)y_v}\,,
\eal\eeq
where $\tilde{k}_j=\pm k_{j,1}\pm\cdots \pm k_{j,m_j}$, and by $k_{j,l}$ we denote all the momenta at the $j$-th vertex. Of course, due to momentum conservation, all of them can be ultimately related in terms of the $n$ external momenta $\vec{k}_i$. The maximum enhancement of the reduced $n$-point function is then clearly achieved for configurations with all of the $\tilde{k}_j$ vanishing. This condition notably requires all the momenta at each vertex to be collinear, and therefore the external momenta must also be collinear. Once again, this collinearity is necessary but not sufficient. Thus the optimal enhancement is attained for the flattened shapes verifying $\tilde{k}_j=0$. We immediately get in that case
\beq
\frac{\langle\zeta^n\rangle}{\langle \zeta^2\rangle^{n-1}}\sim x^{2(m_1+\cdots+m_v)-3v}\,.
\eeq
Noting that for a tree diagram one has $m_1+\cdots+m_v=n+2v-2$, this result thus reproduces the expression quoted in the main text.


\end{document}